\documentclass[allclo]{FBSart}
\usepackage{amsfonts}
\usepackage{amssymb}
\usepackage{epsfig}
\usepackage{bm}
\def\Vec#1{\mbox{\boldmath $#1$}}
\def\eqne{\end{equation}}
 
\def\eqnb{\begin{equation}}

\def\NPB{{Nucl. Phys.} {\bf B}}
\def\PLB{{Phys. Lett.} B}

\def\PRD{{Phys. Rev.} D}

\title{Roles of the color antisymmetric ghost propagator in the infrared QCD}
\author{Sadataka Furui}
\institute{School of Science and Engineering, Teikyo University.\\
1-1 Toyosatodai, Utsunomiya, 320-8551 Japan\thanks{\textit{E-mail address:} furui@umb.teikyo-u.ac.jp }}

\runningauthor{Sadataka Furui}
\runningtitle{Roles of the color antisymmetric ghost propagator in the infrared QCD}
\begin{document}

\maketitle

\begin{abstract}
The results of Coulomb gauge and Landau gauge lattice QCD simulation do not
agree completely with continuum theory. There are indications that the
ghost propagator in the infrared region has strong fluctuation whose modulus
is compatible with that of the color diagonal ghost propagator.
After presenting lattice simulation of configurations produced with Kogut-Susskind fermion (MILC collaboration) and those with domain wall fermion (RBC/UKQCD collaboration), I investigate in triple gluon vertex and the ghost-gluon-ghost vertex how the square of the color antisymmetric ghost contributes.  Then the effect of the vertex correction to the gluon propagator and the
 ghost propagator is investigated. 

Recent Dyson-Schwinger equation analysis suggests the ghost dressing function $G(0)=$ finite and no infrared enhancement or $\alpha_G=0$. But the ghost propagator renormalized by the loop containing a product of color antisymmetric ghost is expected to behave as $< c\bar c>_r =-\frac{G(q^2)}{q^2}$ with $G(q^2)\propto q^{-2\alpha_G}$ with $\alpha_G = 0.5$, if the fixed point scenario is valid.  I interpret the $\alpha_G=0$ solution should contain a vertex correction. The infrared exponent of our lattice Landau gauge gluon propagator of the RBC/UKQCD is $\alpha_D=-0.5$ and that of MILC is about -0.7.

A possible interpretation of the origin of the fluctuation is given.
\end{abstract}


\section{Introduction}
Color confinement and chiral symmetry breaking are the most important characteristics of the infrared (IR) QCD. We studied their mechanism by measuring the gluon propagator $D_A(q^2)$, the ghost propagator $D_G(q^2)$ \cite{FuNa04,FuNa04a,FuNa05a,FuNa06c,FuNa06d,FuNa06} and the quark propagator \cite{FuNa05} in lattice QCD and comparing results with those of Dyson-Schwinger equation (DSE).
From the condition on the IR fixed point of the running coupling in the Landau gauge which is calculated as $\alpha_s(q^2)=q^6 D_G(q^2)^2 D_A(q^2)$, 
the ghost dressing function defined as $G(q^2)=q^2 D_G(q^2)$ in $d-$dimensional system was shown to behave as $G(q^2)\propto q^{-2\alpha_G}$ \cite{Zw02}  and
 $q^2 D_A(q^2)\propto q^{-2\alpha_D}$ with  
\begin{equation}
2\alpha_G+\alpha_D+\frac{4-d}{2}=0\label{GZ}
\end{equation}
in the IR, or it is more singular than the free case.

 Recently an SU(2) gluon propagator of large lattice was shown to be infrared finite\cite{CM07} which was in conflict with the prediction of the DSE\cite{SHA98,AS00} which showed that it is infrared vanishing, or the IR exponents $\alpha_D$ which is defined as $q^2 D_A(q^2)\propto q^{-2\alpha_D}$ is smaller than -1.  
The results could be attributed to the finite size effect, since
infrared vanishing gluon propagator was predicted also from the continuum theory of Zwanziger \cite {Zw91,Zw94,Zw02,Zw07} which considered uniqueness of the gauge field or the celebrated Gribov problem \cite {Gr78} and proposed restriction of gauge configuration to the fundamental modular region, and predicted without
incorporating two-loop contribution $\kappa=\alpha_G=0.59$. 

In the DSE approach, by incorporating the two-loop gluon contribution, an IR finite ($\alpha_D=-1$) gluon propagator was also proposed \cite {Bl03}, although the incorporation of the two-loop or squint diagram in this work is not without ambiguity \cite{Fis06}. 
  Recently a new solution of DSE with IR exponent $\alpha_D=-1$ for the gluon propagator but $\alpha_G=0$ for the ghost propagator was proposed \cite{Paris08, DSVV07}.  In this case, the running coupling in the Landau gauge vanishes at $q=0$.  This behavior was observed in our lattice simulations \cite{FuNa05a} but we concluded that it is an artefact. There are arguments against the IR exponents of the new DSE \cite{AHS08}.

 One should ask the validity of the definition of the running coupling which is based on the tree approximation. I observed in \cite{F08} that the loop corrections through the color antisymmetric ghost could affect IR features of QCD.  The color antisymmetric ghost and its relation to ghost condensates was discussed in \cite{Du05} and the upper limit of the modulus of color antisymmetric ghost is measured in SU(2) \cite{CMM05,CM06} and in SU(3) \cite{FuNa06d,FuNa06}. 

The running coupling  extracted by the JLab group from the experiments shows freezing to a value close to 3.14 \cite{De07}, and the conformal field theory based on the Crewther relation predicts $\alpha_s(0)=\pi$ \cite{BL95,BMMR03}.  A lattice simulation of unquenched configurations in Coulomb gauge is consistent with the JLab extraction \cite{FuNa07}.   

In this paper I summarize our lattice simulation data of the ghost propagator in sect.2 and discuss the effect of color antisymmetric ghost in the gluon-ghost-ghost vertex in sect 3.  Since the color antisymmetric ghost could modify the
ghost-gluon vertex, I discuss a reconsideration of the Slavnov identity due to the presence of the color antisymmetric ghost in sect 4. A discussion and conclusion are given in sect 5. 

\section{Ghost propagators in lattice simulations}

In Landau gauge ($\partial_\mu A_\mu=0$) and in Coulomb gauge ($\partial_i A_i=0$), we adopt the $\log U$ type gauge field $A$, i.e. $U_{x,\mu}=e^{A_{x,\mu}},\ A_{x,\mu}^{\dag}=-A_{x,\mu},$
and for the gauge uniqueness we minimize 
\begin{equation}
F_U(g)=||A^g||^2=\sum_{x,\mu=1,2,3,4}{\rm tr}
 \left({{A^g}_{x,\mu}}^{\dag}A^g_{x,\mu}\right).
\end{equation}
in Landau gauge and
\begin{equation}
F_U[g]=||{\Vec A}^g||^2=\sum_{x,i=1,2,3}{\rm tr}
 \left({{A^g}_{x,i}}^{\dag}A^g_{x,i}\right),
\end{equation}
in the Coulomb gauge.

The gauge field $A_0(x)$ in Coulomb gauge can be fixed by the following minimizing function of $g(x_0)$.
\begin{equation}
F_U[g]=||{A^g}_0||^2=\sum_{x}{\rm tr}
 \left({{A^g}_{x,0}}^{\dag}A^g_{x,0}\right),
\end{equation}
but in this work I leave the remnant gauge unfixed.

In Landau gauge, the ghost propagator is defined as
\begin{eqnarray}
FT[D_G^{ab}(x,y)]&=&FT\langle tr ( \Lambda^a \{({\cal  M}[U])^{-1}\}_{xy}
\Lambda^b  \rangle,\nonumber\\
&=&\delta^{ab}D_G(q^2),  \nonumber
\end{eqnarray}
where ${\mathcal M}=-\partial_\mu D_\mu.$
We solve the equation with plane wave sources.
\begin{equation}
-\partial_\mu D_\mu f_s^b({x})=\frac{1}{\sqrt V}\Lambda^b \sin{q}{x}
\end{equation}
\begin{equation}
-\partial_\mu D_\mu f_c^b({x})=\frac{1}{\sqrt V}\Lambda^b \cos{q}{x}.
\end{equation}

The color diagonal ghost propagator is defined as
\begin{eqnarray}
&&D_G(q)=\frac{1}{N_c^2-1}\frac{1}{V}\nonumber\\
&&\times\delta^{ab}(\langle \Lambda^a\cos{q}{x}|f_c^b({x})\rangle+\langle \Lambda^a\sin{q}{x}|f_s^b({x})\rangle),
\end{eqnarray}
and the color antisymmetric ghost propagator is defined as
\begin{eqnarray}
&&\phi^c(q)=\frac{1}{\mathcal N}\frac{1}{V}\nonumber\\
&&\times f^{abc}(\langle \Lambda^a\cos{q}{x}|f_s^b({x})\rangle-\langle \Lambda^a\sin{q}{x}|f_c^b({x})\rangle)\nonumber\\
\end{eqnarray}
where ${\mathcal N}=2$ for SU(2) and 6 for SU(3).

In our lattice simulation we use the Kogut-Susskind (KS) fermion of MILC collaboration \cite{MILC96,MILC01}  and the domain wall fermion (DWF) of RBC/UKQCD collaboration \cite{QCDOC07}, whose lattice size and parameters are shown in TABLE \ref{config_1}.
 
\begin{table*}
\begin{center}
\caption {The parameters of the lattice configurations.}
\begin{tabular}{c c c c c c c c}
   &$\beta$ &N$_f$ &$m$  &$1/a$(GeV)& $L_s$ & $L_t$ &$a L_s$(fm) \\
\hline
SU(2)      & 2.2 & 0 &         &0.938 & 16 & 16 & 3.37 \\
\hline
SU(3)      & 6.45 & 0 &         &3.664 & 56 & 56 & 3.02\\
\hline
MILC$_{ft1}$ &5.65&  2&  0.008   & 1.716 & 24 &12& 2.76 \\
MILC$_{ft3}$ &5.725&  2& 0.008   & 1.914 & 24 &12& 2.47 \\
MILC$_{ft5}$ &5.85&  2& 0.008   & 2.244 & 24 &12& 2.11 \\
\hline
MILC$_c$ &6.83($\beta_{imp}$)&   2+1   & 0.040/0.050& 1.64 & 20 &64&2.41\\
       &6.76($\beta_{imp}$)&  2+1    & 0.007/0.050& 1.64 & 20 &64&2.41\\
\hline
MILC$_f$ &7.11($\beta_{imp}$)& 2+1     & 0.0124/0.031 & 2.19 &28 & 96&2.52\\
       &7.09($\beta_{imp}$)& 2+1     & 0.0062/0.031 & 2.19 &28 & 96&2.52\\
\hline 
MILC$_{2f}$ &7.20($\beta_{imp}$) &2& 0.020& 1.64 & 20 &64&2.41 \\
\hline
MILC$_{3f}$ &7.18($\beta_{imp}$) &3& 0.031 & 2.19 &28 & 96&2.52 \\
\hline
DWF$_{01}$ &2.13($\beta_{I}$)&2+1 & 0.01/0.04 & 1.743(20) &16 & 32 & 1.81   \\
DWF$_{02}$ &2.13($\beta_{I}$)&2+1 & 0.02/0.04 & 1.703(16) &16 & 32 & 1.85  \\
DWF$_{03}$ &2.13($\beta_{I}$)&2+1 & 0.03/0.04 & 1.662(20) &16 & 32 & 1.90  \\
\hline
\end{tabular}\label{config_1}
\end{center}
\end{table*}

In \cite{FuNa06d,FuNa06}, we showed that the modulus of the color antisymmetric ghost propagator of quenched configuration is small and its variation is large and in \cite{FuNa05} the color diagonal ghost propagator is essentially temperature independent while the color antisymmetric ghost is temperature dependent. 

\begin{figure}[hbt]
\begin{center}
\epsfig{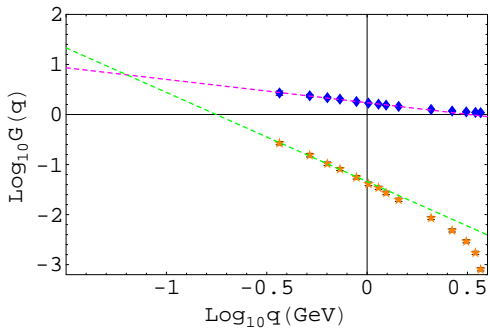}
\caption{The color antisymmetric ghost dressing function $\log_{10}(|\phi({q})|q^2)$ (Orange stars)
 and color diagonal ghost dressing function $\log G(q)$ (Blue diamonds) of quenched SU(2). (67 samples)
Slope defined by the lowest six points of the former is -1.8(1) and that of the latter is -0.45(2).}\label{g_phi_su2}
\end{center}
\end{figure}

\begin{figure}
\begin{center}
\epsfig{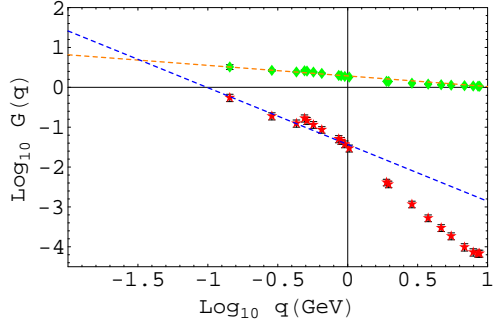}
\caption{The color antisymmetric ghost dressing function $\log_{10}(|\phi({q})|q^2)$ (Red stars)
 and color diagonal ghost dressing function $\log G(q)$ (Green diamonds) of MILC$_f (\beta_{imp}=7.09)$. (21 samples)
Slope defined by the lowest three points of the former is -1.4(3) and that of the latter is -0.27(8).}\label{g_phi_milc709}
\end{center}
\end{figure}

In Figs.\ref{g_phi_su2} and \ref{g_phi_milc709} we show the color diagonal ghost dressing function $G(q^2)$ and the color antisymmetric ghost propagator $\phi(q)$ multiplied by $q^2$ of quenched SU(2) after parallel tempering gauge fixing \cite{FuNa04} and those of unquenched MILC$_{3f}$ configurations, respectively. Errorbars are standard deviations.

\begin{figure}
\begin{center}
\epsfig{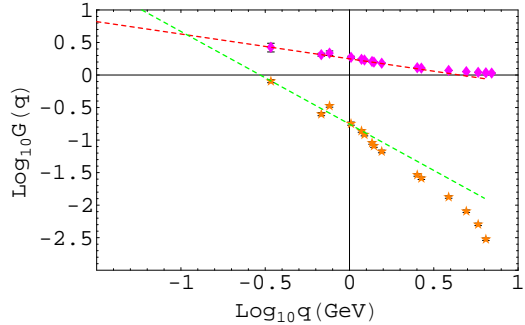}
\caption{The color antisymmetric ghost dressing function $\log_{10}(|\phi({q})|q^2)$ (Orange stars)
 and color diagonal ghost dressing function $\log G(q)$ (Magenta diamonds) of DWF$_{01}$. (50 samples)
Slope defined by the lowest three points of the former is -1.4(1) and that of the latter is -0.38(8).}\label{g_phi_dwf}
\end{center}
\end{figure}

The slope of the color diagonal and that of the modulus of the color antisymmetric ghost depend on the number of flavors. The extrapolated fitting lines of SU(2) and MILC$_{3f}$ cross at around $q=60$MeV and MILC$_f (\beta_{imp}=7.09)$ at around 30MeV (Fig.\ref{g_phi_milc709}). In the case of DWF$_{01}$, they cross at around 100MeV (Fig.\ref{g_phi_dwf}).  Since the lattice size $L$ of DWF$_{01}$ is less than 2 fm, 
finite size effect on the color antisymmetric ghost \cite{CM06} could be important in this case. 
If there are ghost condensates, the color antisymmetric ghost propagator is expected to tend to a constant in the IR. 

In the UV, the color diagonal ghost dressing function remains to be a constant, i.e. $\alpha_G=0$ even when the color antisymmetric
 ghost exists. In \cite{F08}, I showed that color antisymmetric ghosts yield a dominant component of the propagator const$/q^2$ in UV when the IR exponent has $\alpha_G=0.5$. 
In the next section I discuss the gluon-ghost-ghost vertex.

\section{Color antisymmetric ghost in the gluon-ghost-ghost vertex}

\begin{figure}[htb]
\begin{center}
\epsfig{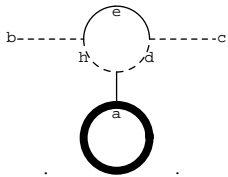}
\caption{The dressing of the ghost propagator by the gluon. The dashed line represents a ghost, the thin line a gluon, and the thick line a quark.}
\label{ghost_dress}
\end{center}
\end{figure}

\begin{figure}
\begin{center}
\epsfig{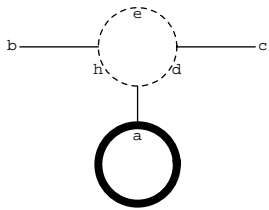}
\caption{The ghost loop contribution in the gluon propagator. The dashed line represents a ghost and the thick line is a quark.}
\label{gluon_dress}
\end{center}
\end{figure}

In \cite{F08}, I showed that in the triple gluon vertex and in the gluon-ghost-ghost vertex, product of color antisymmetric ghost produces real matrix elements when the product makes a color index in Cartan subalgebra.
The vertex can be inserted in the ghost propagator as in Fig.\ref{ghost_dress} or in the gluon propagator as in Fig.\ref{gluon_dress}.


In the ghost propagator, color indices $h$ and $d$ in Fig.\ref{ghost_dress} can make a color antisymmetric ghost when the index $a$ is in Cartan subalgebra, and thus the ghost propagator is not necessarily color diagonal and fluctuates by the operator $f^{abc}$.

On the contrary, in the gluon propagator, color indices $e,d,h$ shown in Fig.\ref{gluon_dress} can be rotated. When the color index $a$ is specified in Cartan subalgebra, color antisymmetric ghost can be chosen not only at $d$ and $h$ but also at $d$ and $e$ or $h$ and $e$. Relative sign of the latter two cases and the
 former is random \cite{F08}.   
Thus the gluon propagator is effectively color diagonal.

\begin{figure}[htb]
\begin{center}
\epsfig{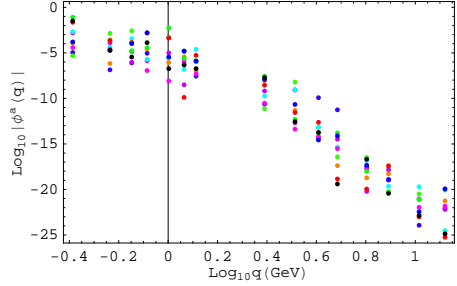}
\caption{Scatter plot of the color antisymmetric ghost dressing function $\log_{10}(|\phi({q})|q^2)$ of quenched $SU(3)$. (10 samples)
}\label{g_phi_quench_su3}
\end{center}
\end{figure}

In lattice simulation, the Kugo-Ojima criterion is satisfied in the unquenched simulation but by about 80\% in the quenched simulation.  We investigated differences of the color SU(3) ghost propagator in Landau gauge and observed that the color antisymmetric ghost propagator in quenched configuration of $56^4$ lattice is random as shown in Fig.\ref{g_phi_quench_su3}, but that in the 
unquenched configuration its randomness is lost and its slope of the modulus as the function of momentum is steeper than that of the color diagonal ghost propagator and its extrapolation to IR becomes compatible with that of the color diagonal component.

The product of the color antisymmetric ghost affects the triple gluon or
ghost-gluon-ghost vertices and gluon propagator and the ghost propagator via
loop corrections. The QCD effective coupling is affected by the ghost propagator and there appear differences in Coulomb gauge and in Landau gauge.

In \cite{F08}, I found that the ghost propagator with a gluon-ghost-ghost loop
with a tadpole vertex between the two ghost propagators introduces an effective
propagator of a const$/p^2$ if the IR exponent of the ghost propagator
$\alpha_G=0.5$. It means that in the UV, the dominant component of the
ghost propagator is const$/p^2$ as observed recently by the Dyson-Schwinger
approach \cite{Paris08, DSVV07}. In the Dyson-Schwinger approach of \cite{AS00}
the exponent in the UV is the same as that in the IR.
Our calculation of the loop diagram and the lattice simulation of the ghost 
propagator suggest that the exponent in the UV is 0 but that in the IR
is finite and enhances the singularity. If $\alpha_G=0$, additional vertex
singularity of $q^{-4\kappa}=q^{-2}$ yields the same const/$q^2$ dominant behavior in UV.

\section{A comparison with continuum theory of infrared QCD}

The origin of the fluctuation of the ghost propagator could be the Gribov copy
effect and I would like to discuss applicability of the Faddeev-Popov quantization in the infrared region.

In the Faddeev-Popov quantization method, Zavialov showed in his book \cite{Za90} 
that for an arbitrary functional $F(A,\bar c,c)$ depending on fields $A_\mu^a(x_i), \bar c^a(y_j), c^a(z_k)$ and let $\delta F(A,\bar c,c)$ denotes its variation under BRST transformation, the Green function of $\delta F$ is zero, i.e. $\langle S\delta F\rangle=0$. 

Zavialov took $F=A_\mu^a(x)\bar c^d(y)$ and calculated $\langle S\delta F\rangle=\langle \delta F S\rangle$.
He showed that
\begin{eqnarray}
&&F(-i\square ^{\mu\nu}\hat A_\nu(x))=\frac{\delta F}{\delta A^a_\mu(x)},
\end{eqnarray}
which implies that $\hat A_\mu(x)$ can be replaced by the functional derivative,
 and would be valid since gluon propagator is color diagonal.

In the case of ghost, however it is not evident that the ghost field
can be replaced by the functional derivative as
\begin{eqnarray}
F(-i\square\hat c^a(x))=\frac{\delta F}{\delta \bar c^a(x)}.
\end{eqnarray}

 Since gauge fixing can be defined only locally, gauge potentials are partitioned into patches corresponding to copies. The Faddeev-Popov quantization in the high energy region could be 
inefficient in infrared due to overlapping of patches.  There are proposals of stochastic quantization \cite{Zw02,HK98,HK00,ZJ01}, but whether the unique gauge is
attained through the stochastic quantization is not evident due to large fluctuation from color antisymmetric ghost. 
 
The argument does not mean the Slavnov identity is broken, since in the tree level, the color antisymmetric contribution cancels out among themselves.
However, in the one loop, products of the color antisymmetric ghosts affect IR features of QCD. 

Similar argument applies to the renormalization of the gauge theory.
In 1971 t'Hooft showed that in the massless Yang-Mills field theory, the gauge invariance can be restored by incorporating finite number of counter terms including ghosts and longitudinal gluons in the system \cite{tH71}.
He showed, if the longitudinal gluon forms a pair of color diagonal ghosts and the loop intersects with the circle that denotes a set of particles on mass-shell, the contribution is cancelled by the cutoff denoted by $\Lambda$ (Fig.7). 
But when a pair of color antisymmetric ghosts are produced and there are Gribov copies, whether one can define $\Lambda$ globally is not clear.  
In compact lattices, we observe fluctuation of propagators is large for momenta  which deviate from the diagonal of the four dimensional system (cylinder cut region), which may be related to this issue.

\begin{figure}
\begin{center}
\epsfig{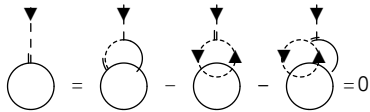}
\caption{The cancelling mechanism of longitudinal gluon and ghosts. }
\end{center}
\end{figure}

In a semiperturbative DSE calculation \cite{SMWA05}, and in
a lattice simulation \cite{CMM08}, the infrared ghost-gluon vertex in Landau gauge was claimed to be close to that of tree level. These analyses are done in
quenched approximation, where modulus of color antisymmetric ghost is random
as shown in Fig.\ref{g_phi_quench_su3}. It is not clear whether, in unquenched systems, a product of color antisymmetric ghosts that appear in one loop level does not affect the theory. 

\subsection{Kugo-Ojima color confinement criterion}

In 1971, Taylor \cite{Ta71} pointed out that the Ward identity in QED i.e. $Z_1/Z_2=1$ can be extended to QCD as $Z_1/Z_3=Z_{\bar\psi \psi A}/Z_\psi$.
In 1979 Kugo and Ojima \cite{KO79} showed that if color symmetry is not broken
$Z_1/Z_3=\tilde Z_1/\tilde Z_3=Z_{\bar\psi \psi A}/Z_\psi$.
The IR exponent of the ghost changes the color confinement criterion of
Kugo and Ojima, which says that in the Landau gauge a coefficient $u^{ab}(q^2)=\delta^{ab}u(p^2)$ and if at momentum zero $u(0)=-1$ the color confinement occurs. 
 
\begin{eqnarray}
&&(\delta_{\mu\nu}-{q_\mu q_\nu\over q^2})u^{ab}(q^2)\nonumber\\
&&={1\over V}
\sum_{x,y} e^{-ip(x-y)}\langle  {\rm tr}\left({\Lambda^a}^{\dag}
D_\mu \displaystyle{1\over -\partial D}[A_\nu,\Lambda^b] \right)_{xy}\rangle.\nonumber
\end{eqnarray}

When we define
\[
\langle c\bar c\rangle\equiv -\frac{1}{q^2 {\mathcal G}(q^2)},
\]
\[
\langle (A_\mu\times c)\bar c\rangle_{1PI}\equiv -iq_\mu {\mathcal F}(q^2)
\]
and 
\begin{equation}
\langle D_\mu c\bar c\rangle=\langle \partial_\mu c\bar c\rangle+\langle (A_\mu\times c)\bar c\rangle\equiv iq_\mu(1+{\mathcal F}(q^2))\frac{1}{q^2 {\mathcal G}(q^2)},\label{KO79}
\end{equation}
we obtain, provided $c$ in $A_\mu\times c$ and $\bar c$ in $A_\nu\times \bar c$ do not couple color diagonal,
\begin{eqnarray}
\langle D_\mu c(A_\mu\times \bar c)\rangle&=&\langle \partial_\mu c\bar c\rangle\langle c(A_\mu\times \bar c)\rangle_{1PI}
+\langle (A_\mu\times c)(A_\nu\times \bar c)\rangle_{1PI}\nonumber\\
&\ne& (\delta_{\rho\mu}-\frac{q_\mu q_\rho}{q^2})\langle c(A_\mu\times \bar c)\rangle_{1PI}\label{1pieq}.
\end{eqnarray}

Although our finite size lattice simulation suggests ${\mathcal G}(0)=0$, there could be a contribution of the ghost propagator of the type appearing  between $c$ and $\bar c$ that is not proportional to $\delta^{ab}$.  Thus, $1+u(0)=1+{\mathcal F}(0)$ is not necessarily equal to 0 and ${\mathcal F}(0)$ can deviate from -1.  

In the continuum limit, a solution with ${\mathcal G}(0)=$finite and $\alpha_G=0$ is possible \cite{Paris08}. 
Then $1+{\mathcal F}(0)$ is not necessarily of higher order 0, as required by Kugo \cite{KUG}. 

\subsection{QCD effective coupling}

\begin{figure}
\begin{center}
\epsfig{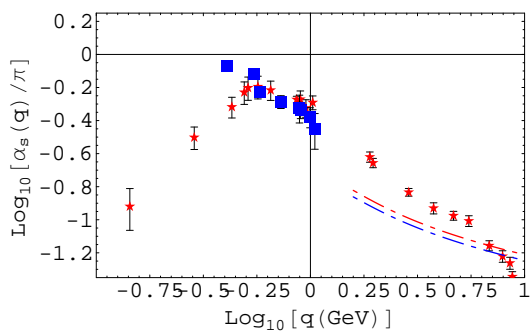}
\caption{The running coupling $\alpha_s(q)/\pi$ of MILC$_f$ in Landau gauge  The pQCD result of $N_f=3$ (upper dash-dotted line) and $N_f=2$ (lower dashed line) and the extraction of JLab are also plotted(blue boxes).}\label{alp_landau}
\end{center}
\end{figure}

In Landau gauge, the running coupling $\alpha_s({q}^2)$ can be calculated
from the ghost-gluon coupling, the triple gluon coupling or the quark-gluon couping.
The ghost-gluon coupling is given by the product of the ghost dressing function squared times the gluon dressing function. 
\begin{equation}
  q^6 D_G(q)^2 D_A(q)\propto \alpha_s(q^2).
\end{equation}
The small ghost dressing function causes IR suppression of the running coupling, as shown in Fig.\ref{alp_landau}.

\begin{figure}
\begin{center}
\epsfig{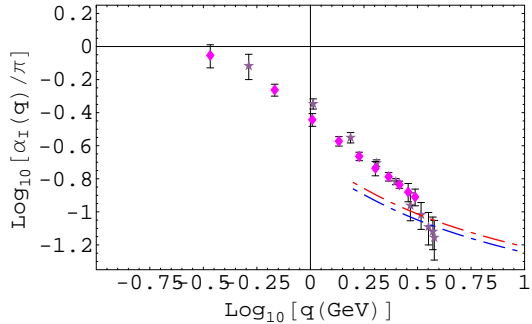}
\caption{The running coupling $\alpha_I(q)/\pi$ of MILC$_{3f}$ and MILC$_{2f}$ in Coulomb gauge.  The pQCD result of $N_f=3$ (upper dash-dotted line) and $N_f=2$ (lower dashed line) are also plotted. 
}\label{alp_coulomb}
\end{center}
\end{figure}

 In Coulomb gauge, the corresponding coupling constant $\alpha_I({\Vec q}^2)$ 
is obtained by choosing the interpolation gauge parametrized by $\eta$ between the Landau gauge ($\eta=1$) and the Coulomb gauge ($\eta=0$). Using the
limit of an integration over the 4th component of the momentum $\displaystyle \lim_{\eta\to 0}\int_{-\infty}^\infty \frac{dq_0}{2\pi}\frac{\eta|\Vec q|^5}{({\Vec q}^2+\eta^2 {q_0}^2)^3}=\frac{3}{16}$, we obtain
\begin{equation}
  {\Vec q}^5 D_G({\Vec q})^2 D_A^{tr}({\Vec q})\propto \alpha_I({\Vec q}^2),
\end{equation}
where $D_A^{tr}({\Vec q})$ is the 3-dimensional gluon propagator\cite{FZ05}. 
 No IR suppression occurs in $\alpha_I({\Vec q})$, as shown in Fig.\ref{alp_coulomb}.

The fixed point scenario suggests $\alpha_D+2\alpha_G=0$. A comparison of this conditions with lattice data is shown in TABLE \ref{exponent}.  We observe that the lattice data deviate from the theory by about 20\%.
 The Paris and Gent solution \cite{Paris08,DSVV07} $\alpha_D=-1, \alpha_G=0$ would mean that one needs correction of the vertex renormalization. The Graz solution \cite{AHS08} $\alpha_D=-1.2, \alpha_G=0.6$ is not compatible with the gluon propagator of large lattice \cite{CM07}.  Whether the gluon propagator is affected by the two-loop diagram including color antisymmetric ghosts \cite{F08} needs to be investigated. I remark that when a large instanton is present, fermions becomes massive due to chiral symmetry breaking and they decouple from gluons \cite{AC75}.  In this case, the fixed point scenario reflects only an approximate feature of IR-QCD, and it could be violated. 

\begin{table}
\caption{Infrared exponents of MILC and DWF configurations in Landau gauge.}
\begin{center}
\begin{tabular}{c c c c c c}
&$\beta_{imp/I}$ &$\alpha_G$ & $\alpha_D$& $\alpha_D+2\alpha_G$ &$3\alpha_D-2\alpha_G$\\
\hline
MILC$_c$& $ 6.76$ &0.25	&-0.60	&-0.10 & -2.30(20)\\
@	& $ 6.83$ &0.23	&-0.57	&-0.11 & -2.17(20)\\
MILC$_f$& $ 7.09$ &0.24	&-0.67	&-0.19 & -2.49(20)\\
@	& $ 7.11$ &0.23	&-0.65	&-0.19 & -2.41(20)\\
\hline
DWF$_{01}$ & $2.13$ & 0.19 & -0.49 & -0.11 & -1.85(20)\\
DWF$_{02}$ & $2.13$ & 0.16 & -0.53 & -0.21 & -1.94(20)\\
DWF$_{03}$ & $2.13$ & 0.17 & -0.47 & -0.13 & -1.76(20)\\
\hline 
\end{tabular}
\end{center}
\label{exponent}
\end{table}

\section{Conclusion and discussion}

The color antisymmetric ghost introduces a fluctuation of propagators in the one loop level. A comparison of the quenched and the unquenched simulation of SU(3) Yang-Mills theory suggests that a quark has the effect of magnifying the square norm of the color antisymmetric ghost propagator and reduces its fluctuation.
The asymptotic behavior of the color diagonal ghost dressing function and $q^2$
times the color antisymmetric ghost in the IR is not known. When the crossing point of the extrapolation of the color diagonal ghost dressing function and $q^2$ times the color antisymmetric ghost is relatively large ($\simeq 60$MeV), as in the case of DWF, the color diagonal ghost propagator in IR becomes unstable.

 The gluon-ghost-ghost loop contribution with the IR exponent $\alpha_G=0.5$ produces the ghost propagator with the dominant component of $\alpha_G=0$ in UV.  It is compatible with the recent findings of DSE equation.  Historically high
infrared singularity of the ghost propagator was required to cancel the
$q^{-4}$ IR singularity of gluon propagator in the tree level \cite{Man79}. 
Recently in 3-dimensional Landau gauge SU(2) Yang-Mills theory, infrared divergence of the three gluon vertex was suggested \cite{CMM08}. It is in conflict
with the 4-dimensional Landau gauge SU(3) Yang-Mills theory \cite{Orsay02},
which asserts that the three gluon vertex is infrared vanishing due to instanton effects. If instantons play a role in infrared QCD, quenched and unquenched simulation would produce differences, since fermionic zero mode which is absent in quenched simulation, is expected to cancel bosonic zero mode divergence \cite{F08b}. 

The IR exponent $\alpha_G=0$ and $\alpha_D=-1$ violates the fixed point scenario of the running coupling which requires $\alpha_D+2\alpha_G=0$.  The difference of the ghost-gluon coupling of the Coulomb gauge and that of the Landau gauge below $q\sim$ 0.6 GeV suggests the vertex renormalization, i.e. the IR suppression of $\alpha_s(q)$ in the Landau gauge may be due to the singularity of the color antisymmetric ghost propagator that disturbs the color diagonal part and invalidates the tree level approximation. The agreement of the quark-gluon coupling of the Coulomb gauge, that of the Landau gauge and the JLab experimental data \cite{F08b} suggest that the different momentum dependence of Landau gauge effective quark-gluon coupling and the effective ghost-gluon coupling in the IR is due to our incomplete treatment of the ghost propagator, since I think the universality should be preserved in the whole momentum region.

 Since lattice data are taken on the torus and suffers from finite size effects, we need to continue careful comparisons of ans\"atze and results of the DSE and the lattice calculation.
More extensive study of the Coulomb gauge DWF quark propagator\cite{F08a} of larger lattice is left in the future.

\begin{acknowledge}
A part of this work was done at the Department of Theoretical Physics of University of Graz in March 2008. The author thanks Reinhard Alkofer and Kai Schwenzer for helpful discussions and hospitality, organizer of the workshop on {\it Quarks and Hadrons in strong QCD} (St. Goar) for the support and providing me the oppotunity of discussing with several participants, and Hideo Nakajima for the collaboration in the early stage of this research.
The numerical simulation was performed on Hitachi-SR11000 at High Energy Accelerator Research Organization(KEK) under a support of its Large Scale Simulation Program (No.07-04), on NEC-SX8 at Yukawa institute of theoretical physics of Kyoto University and at Cybermedia Center of Osaka university.
\end{acknowledge}

\end{document}